\documentclass[11pt]{article}
\usepackage{amsxtra}
\usepackage{lipsum}
\usepackage{braket}
\usepackage{amsmath, amssymb, mathrsfs}
\usepackage[T1]{fontenc}
\usepackage[utf8]{inputenc}
\usepackage{graphicx,float,color}
\usepackage{subfigure}
\usepackage{subcaption}
\usepackage{changepage}
\usepackage{soul}

\title{Spontaneous baryosynthesis with large initial phase}
\author{M.A. Krasnov \thanks{Speaker. This is the write-up of the talk, based on Ref.~\cite{Krasnov:2025fbo}, presented by M. A. Krasnov at the 28th Bled Workshop, "What Comes Beyond the Standard Models", Bled, Slovenia, July 6 - 17, 2025.}\\ Research Institute of Physics Southern Federal University,\\ Rostov-on-Don, 344090, Russia \\ M.Yu. Khlopov \\ Virtual Institute of Astroparticle Physics, Paris, 75018, France \\ U. Aydemir \\ Department of Physics, Middle East Technical University,\\ Ankara, 06800, Türkiye}

\begin{document}
\maketitle

\begin{abstract}
We numerically investigate particle production by a pseudo-Nambu-Goldstone boson (pNGB) in spontaneous baryogenesis, focusing on large initial misalignment angles. Our analysis confirms the established cubic dependence of the baryon asymmetry on the initial phase for small angles. However, this scaling breaks down for larger angles, with particle production saturating as the initial phase approaches $\pi$ in Minkowski spacetime. 
\end{abstract}

\section{Introduction}
\label{introduction}

Observational data unequivocally confirms the existence of a universe dominated by matter, with a significant asymmetry between baryons and antibaryons. This is puzzling, as fundamental physics offers no obvious reason for such an imbalance in the production of particles and antiparticles. This baryon asymmetry is quantified by the present-day baryon-to-entropy ratio, $(\Delta n_{\mathrm{B}}/s)_0 \simeq 8.6\times 10^{-11}$ ~\cite{Planck:2018vyg}. For decades, a major challenge in cosmology has been to identify a physical process that naturally explains this value, rather than simply treating it as an initial condition of the universe. {\color{black} The foundational framework for this, proposed by Sakharov and Kuzmin \cite{Sakharov:1967dj, KUZMIN}, connects the generation of a baryon excess from an initially symmetric state to CP-violating processes that occur out of equilibrium and that do not conserve baryon number. Subsequent research has expanded this idea, leading to various proposed mechanisms that tie the origin of the baryon asymmetry to new physics beyond the Standard Model.}

One such mechanism, known as spontaneous baryogenesis, was introduced in Refs.~\cite{Cohen:1987vi,Cohen:1988kt} and further explored in Refs.~\cite{Dolgov_1995,Dolgov_1997}  
In this scenario, the asymmetry arises from the relaxation of a (pseudo) Nambu-Goldstone boson specifically, the phase $\theta=\phi/f$ of a spontaneously broken global $U(1)$ baryonic symmetry toward the minimum of its potential. Here, $f/\sqrt{2}$ corresponds to the magnitude of the vacuum expectation value of the complex scalar field responsible for the symmetry breaking. This field acts as a spectator during inflation, coexisting with the inflaton. An explicit symmetry-breaking term, given by the potential $V(\theta)=\Lambda^4(1-\cos\theta)$\footnote{This is analogous to the QCD axion potential, though here it is not generated by QCD instanton effects.}, tilts the potential and gives mass to the originally massless boson. The field $\theta$ is coupled derivatively to a non-conserved baryonic current via the dimension-5 operator $\mathcal{L}_{\mathrm{B}}=f^{-1}J_{\mathrm{B}}^{\mu}\partial_\mu \phi$, where $J_{\mathrm{B}}^{\mu}= \overline{Q}\gamma^\mu Q$ and $Q$ is a new heavy fermion carrying baryon number. As $\theta$ undergoes damped oscillations, it is converted into either baryons or antibaryons, depending on the direction in which it rolls toward the minimum of the tilted potential. The resulting asymmetry is thus determined by the initial angle $\theta_i $.

This work investigates the consequences of large initial misalignment angles within the spontaneous baryogenesis framework. While the small-angle approximation frequently used in the literature is convenient and insightful, the phase distribution at the end of inflation does not necessarily favor such small values. It is therefore essential to explore the implications of large misalignment angles. The most intriguing starting point is $\theta_i\simeq\pi$, which corresponds to the local maximum of the potential. The phase will then roll down to a minimum at either $\theta=0$ or $\theta=2\pi$, depending on the direction of motion.

Consequently, $\theta_i=\pi$ represents a domain wall separating two degenerate vacuum states. In our analysis, we therefore initiate the motion from $\theta_i\simeq\pi$ to study its impact on baryon asymmetry generation.

While it is possible that inflation is driven by the Nambu-Goldstone boson itself a model known as "natural inflation"~\cite{Freese:1990rb} recent analyses strongly disfavor this scenario~\cite{Alam_2024, dos_Santos_2024} due to tensions with PLANCK data \cite{Planck:2018vyg}, particularly the constraints on the tensor-to-scalar ratio $r$ and the scalar spectral index $n_s$.

In this paper, we assume the Nambu-Goldstone boson responsible for baryogenesis is a spectator field during inflation and remain neutral regarding the specific mechanism driving inflation. \textcolor{black}{We posit that the Nambu-Goldstone boson emerges during inflation, but its classical dynamics are frozen, with only quantum fluctuations being active.}

The structure of this paper is as follows. Section~\ref{model} provides a concise overview of the model that gives rise to the (pseudo) Nambu-Goldstone boson. Section~\ref{distribution} examines the probability distribution of the baryon asymmetry. In Section~\ref{stationary}, we detail our numerical approach to solving the equation of motion in Minkowski space-time.  Our analysis culminates in Section~\ref{baryonasymmetry} with the computation of the baryon asymmetry, where we illustrate its dependence on the Nambu-Goldstone boson's initial value. We conclude with a summary and discussion. Throughout this work, we use units where $c=\hbar=k_B=1$, unless stated otherwise.

\section{Theoretical Framework\label{model}}

We begin by outlining the fundamentals of the spontaneous baryogenesis model, based on the seminal works of A. Dolgov and colleagues~\cite{Dolgov_1995, Dolgov_1997}. The central element is a complex scalar field $\Phi$ that experiences spontaneous symmetry breaking, producing a Nambu-Goldstone boson which subsequently facilitates baryon number generation.
The Lagrangian includes $\Phi$ along with heavy fermionic fields: a fermion $Q$, postulated to carry baryon charge, and a lepton field $L$:
\begin{multline}
\label{UnbrokenL}
\mathcal{L}=\partial_\mu \Phi^* \partial^\mu \Phi - V(\Phi)+i\overline{Q}\gamma^\mu \partial_\mu Q+i\overline{L}\gamma^\mu \partial_\mu L -\\ - m_Q \overline{Q}Q -
m_L \overline{L}L + g(\Phi \overline{Q} L+ \Phi^* \overline{L}Q).
\end{multline}
The Yukawa interaction term, $g(\Phi \overline{Q} L + \Phi^* \overline{L}Q)$, is critical, as it later enables the production of the $Q$ field and the violation of baryon number.
This Lagrangian is invariant under a classical global $U(1)$ symmetry associated with baryon number, under which the fields transform as:
\begin{equation}
\label{baryonsymmetry}
\Phi \rightarrow e^{i\alpha}\Phi, \quad Q \rightarrow e^{i\alpha}Q, \quad L \rightarrow L.
\end{equation}
The scalar potential $V(\Phi)$ is designed to induce spontaneous symmetry breaking (SSB) of this $U(1)$ at the energy scale $f$:
\begin{equation}
V(\Phi)=\lambda \left(\Phi^*\Phi - f^2/2\right)^2.
\end{equation}
This potential generates a nonzero vacuum expectation value (VEV), $\langle\Phi\rangle = \frac{f}{\sqrt{2}}e^{i\phi/f}$, breaking the $U(1)$ symmetry. Expanding around this VEV reveals the angular degree of freedom $\phi$ as the massless Nambu-Goldstone boson.

Expressing the field as $\Phi(x) = \frac{f}{\sqrt{2}} e^{i\theta(x)}$, where $\theta(x) \equiv \phi(x)/f$, and substituting into the original Lagrangian yields the effective theory below the SSB scale:
\begin{multline}\label{AftSymBroken2}
\mathcal{L}=\frac{f^2}{2}\partial_\mu \theta \partial^\mu \theta + i\overline{Q}\gamma^\mu \partial_\mu Q + i\overline{L}\gamma^\mu \partial_\mu L - m_Q \overline{Q}Q - m_L \overline{L}L +\\+
\frac{gf}{\sqrt{2}}\left(\overline{Q} L e^{i\theta} + \overline{L}Q e^{-i\theta}\right) - V(\theta).
\end{multline}
This Lagrangian remains invariant under the shifted $U(1)$ transformation:
\begin{equation}\label{transformations}
Q\rightarrow e^{i\alpha}Q, \quad L \rightarrow L, \quad \theta \rightarrow \theta + \alpha.
\end{equation}
To generate a mass for the $\theta$ field and provide a potential for it to evolve, an explicit symmetry-breaking term is introduced. This potential, analogous to the axion potential from QCD instantons but treated here as a generic low-energy effect parameterized by a scale $\Lambda \ll f$, is:
\begin{equation}
\label{cosinepotential}
V(\theta) = \Lambda^4(1 - \cos\theta).
\end{equation}
This potential tilts the initial Mexican hat, endowing the pseudo-Nambu-Goldstone boson with a mass $m_\theta \sim \Lambda^2/f$.

{\color{black} The Lagrangian in Eq.~(\ref{AftSymBroken2}) can be rewritten by applying the field redefinition $Q\rightarrow e^{-i\theta(x)}Q$}. {\color{black}This transformation eliminates the phase from the Yukawa interaction and gives rise to a derivative coupling term:
\begin{multline}\label{AftSymBroken3}
\mathcal{L}=\frac{f^2}{2}\partial_\mu \theta \partial^\mu \theta + i\overline{Q}\gamma^\mu \partial_\mu Q + i\overline{L}\gamma^\mu \partial_\mu L - m_Q \overline{Q}Q - m_L \overline{L}L +\\+
\frac{gf}{\sqrt{2}}(\overline{Q} L + \overline{L}Q) + \partial_\mu\theta \overline{Q}\gamma^\mu Q - V(\theta).
\end{multline}

The term $\partial_\mu\theta \overline{Q}\gamma^\mu Q$ is the distinctive feature of spontaneous baryogenesis.\footnote{Ref.~\cite{Cline:2025yvg} raises the concern that the generation of
baryon number is accompanied by an equal lepton-number asymmetry, so that
$B-L=0$ and the resulting asymmetry is subject to sphaleron washout. As stated
in the original articles~\cite{Dolgov_1995,Dolgov_1997}, and as detailed in
the revised version of Ref.~\cite{Krasnov:2025fbo}, this conclusion does not
apply to the charge assignment used here, since the heavy field $L$ does not
carry SM lepton number by construction.}

\section{Asymmetry Distribution\label{distribution}}

\textcolor{black}{
The initial value of the phase field $\theta_i$ at the onset of its oscillations is not fixed but is determined by quantum fluctuations during cosmological inflation. We examine the probability distribution $f(\phi, t)$ for a light scalar field $\phi$ (with $\theta = \phi/f$) during inflation. This distribution can be derived from the Fokker-Planck equation \cite{LINDE1982335,  Vennin_2015}, which, for a massless field ($m \ll H_\star$), results in a Gaussian distribution. Starting from an initial value $\phi_u$ when inflation begins, the probability density of finding the field at value $\phi$ after time $t$ is given as \cite{Belotsky:2018wph}:
\begin{equation}\label{probDensity}
f(\phi,t) = \frac{1}{\sqrt{2\pi}\sigma(t)} \exp\left(-\frac{(\phi - \phi_u)^2}{2\sigma^2(t)}\right),
\end{equation}
where $\sigma(t) = \frac{H_\star}{2\pi} \sqrt{H_\star t}$. This describes the field's random walk due to quantum fluctuations superimposed on the classical slow-roll motion.
The baryon asymmetry produced in spontaneous baryogenesis is highly dependent on the initial phase $\theta_i$ at the end of inflation. Converting the distribution for $\phi$ into one for the phase $\theta_i = \phi_i / f$, and assuming inflation lasts for $N \approx 60$ e-folds ($t \approx 60 H_\star^{-1}$), one obtains the probability distribution for the initial misalignment angle after inflation:
\begin{equation}\label{distr}
f(\theta_i) = \frac{1}{\sqrt{2\pi}\sigma'} \exp\left(-\frac{(\theta_i - \theta_u)^2}{2\sigma'^2}\right),
\end{equation}
where $\sigma' = \frac{H_\star}{2\pi f} \sqrt{60}$.
A key aspect of cosmological inflation is that causally disconnected regions evolve independently. The entire observable universe today originates from approximately $e^{3N} \approx e^{180}$ such independent Hubble patches at the end of inflation. Within each patch, $\theta_i$ is nearly uniform but varies randomly between patches according to the distribution \eqref{distr}. This renders spontaneous baryogenesis an \textit{inhomogeneous} process on super-Hubble scales at this epoch; different regions will yield different baryon asymmetries.
The probability that the misalignment angle in a given Hubble patch deviates from its mean value $\theta_u$ by more than $\pi$ is given by the two-sided tail of a Gaussian distribution~\cite{Belotsky:2018wph}:
\begin{equation}
P\!\left(|\theta_i-\theta_u|>\pi\right)
= \operatorname{erfc}\!\left(\frac{\pi}{\sqrt{2}\,\sigma'}\right)
= 1-\operatorname{erf}\!\left(\frac{\pi}{\sqrt{2}\,\sigma'}\right),
\end{equation}
where $\operatorname{erf}$ is the (Gauss) error function and $\operatorname{erfc}$ is the complementary error function, $\operatorname{erfc}(z)\equiv 1-\operatorname{erf}(z)$.
}

Assuming that the symmetry breaking scale $f$ is similar to the Hubble scale during inflation ($f \approx H_\star$), we find $\sigma' \approx \sqrt{60}/(2\pi) \approx 1.23$, and thus:
\begin{equation}
P(|\theta_i - \theta_u| > \pi) \approx 1 - \text{erf}(\pi) \approx 10^{-5}.
\end{equation}
Although this probability for a single patch is low, the total number of patches is immense. The expected number of patches within our observable universe that have experienced such a large fluctuation is:
\begin{equation}
n_{\text{regions}} = e^{180} \times P(|\theta_i - \theta_u| > \pi) \approx 10^{78} \times 10^{-5} \gg 1.
\end{equation}
Therefore, it is statistically certain that regions with $\theta_i \sim \pi$ exist within our current horizon. This requires a thorough investigation of the baryogenesis mechanism for these large initial misalignment angles, which is the principal objective of this study.
}

\section{Numerical Solution in a Static Universe\label{stationary}}
This section examines the equation of motion in Minkowski space-time, with the simplification of massless fermions. For an arbitrary initial phase, the relevant semiclassical equation of motion is~\cite{Dolgov_1995}:
\begin{multline}\label{start}
\Ddot{\theta}+\cfrac{\Lambda^4}{f^2}\sin{\theta} = -\cfrac{4g^2}{\pi^2} \int_0^\infty \omega^2 d\omega \times \\ \times \int_{-\infty}^0 dt'\sin{(2\omega t')}\sin{[\theta(t+t')-\theta(t)]},
\end{multline}
which can be reformulated as:
\begin{multline}\label{full_equation}
\Ddot{\theta}+\cfrac{\Lambda^4}{f^2}\sin{\theta} = -\cfrac{g^2}{2\pi^2} \lim_{\omega \rightarrow \infty} \int_{-\infty}^0 dt' \left[ \cfrac{\cos{2\omega t'}-1}{t'}\right] \times\\ \times \left[ \Ddot{\theta}(t+t')\cos{\Delta \theta} - \Dot{\theta}^2(t+t') \sin{\Delta \theta} \right],
\end{multline}
where $\Delta \theta = \theta(t+t') - \theta(t)$. It is important to note that Eq.~\eqref{full_equation} is derived from a treatment where the scalar field $\theta$ is classical, while the fermion fields $Q$ and $L$ are treated quantum mechanically. This imposes limitations on the allowed initial conditions for $\theta$. For example, the configuration $\theta_i = \pi$ with $\Dot{\theta}_i = 0$ is not physically meaningful, as it would yield the static solution $\theta= \pi$.

We begin the solution process by rewriting Eq.~\eqref{full_equation} and denoting the integral as:
\begin{multline}\label{MinkowskiSol}
\Ddot{\theta}+\cfrac{\Lambda^4}{f^2}\sin{\theta}
=\cfrac{g^2}{\pi^2} \lim_{\omega \rightarrow \infty} \int_{-\infty}^0 dt' \left[ \cfrac{\sin^2{\omega t'}}{t'}\right]\times \\ \times \left[ \Ddot{\theta}(t+t')\cos{\Delta \theta} - \Dot{\theta}^2(t+t') \sin{\Delta \theta} \right] \equiv \mathcal{I}.
\end{multline}
A crucial step in our approach is to treat $\omega$ as large but finite, effectively introducing a cutoff to the integration limit in \eqref{start}. Since the pseudo-Nambu-Goldstone boson description emerges at energies below $f$, it is physically justified to set the effective theory's cutoff energy at $\omega\sim f$.\footnote{Ref.~\cite{Cline:2025yvg} characterizes the introduction of the
cutoff scale as ad hoc. In the EFT interpretation, however, this scale is not
an additional assumption tied specifically to the memory integral. The radial
mode has been integrated out, and the angular mode remains as the only
low-energy degree of freedom below the symmetry-breaking scale. This is also
reflected in Eq.~(\ref{AftSymBroken3}): writing the canonically normalized
pNGB field as $\phi=f\theta$, the derivative coupling takes the form
$(\partial_\mu\phi/f)\,\overline Q\gamma^\mu Q$. The interactions of the
canonically normalized pNGB field are therefore suppressed by the
symmetry-breaking scale.} Given that the cosine potential becomes significant at scales much lower than $f$ (as indicated before Eq.~(\ref{cosinepotential})), we also have $m=\Lambda^2/f\ll\omega \sim f$.
                      
We now proceed without the limit operator and analyze the integral:
\begin{multline}
\mathcal{I}(t)= \cfrac{g^2}{\pi^2} \int_{-\infty}^0 dt' \left[ \cfrac{\sin^2{\omega t'}}{t'}\right]\times\\ \times \left[ \Ddot{\theta}(t+t')\cos{\Delta \theta} - \Dot{\theta}^2(t+t') \sin{\Delta \theta} \right] .
\end{multline}

Integrating this expression by parts yields:
\begin{multline}
\mathcal{I}(t)=\cfrac{g^2}{\pi^2}\cfrac{\sin^2{\omega t'}}{t'}\Dot{\theta}(t+t')\cdot\cos{[\Delta\theta]}|_{-\infty}^0- \\ -\cfrac{g^2}{\pi^2}\int_{-\infty}^0 dt' \Dot{\theta}(t+t')\cdot\cos{[\theta(t+t')-\theta(t)]} \times \\ \times\left( \cfrac{\omega \sin{(2\omega t')}}{t'}-\cfrac{\sin^2{(\omega t')}}{t'^2} \right).
\end{multline}

Recalling standard representations of the Dirac delta-function:
\begin{equation}
\label{approximation}
\delta(t)=\lim_{\omega \rightarrow \infty} \cfrac{\sin{\omega t}}{\pi t},\,\delta(t) = \lim_{\omega \rightarrow \infty} \cfrac{\sin^2{\omega t}}{\pi \omega t^2}.
\end{equation}

It is clear that the $\omega\to\infty$ limit is problematic and requires
regularization. For numerical calculations, it is therefore sensible to
introduce a physical cutoff. As discussed above, a natural choice is the
characteristic scale of the theory, $f$, below which the Nambu--Goldstone
boson emerges and the effective description in terms of the angular mode is
valid. Taking $\Lambda^2/f\ll\omega$ and assuming that the limiting procedure
captures the leading local dissipative structure, we employ the
approximation
\begin{equation}
\frac{\omega \sin(2\omega t')}{t'}
-
\frac{\sin^2(\omega t')}{t'^2}
\approx
\pi \omega \,\delta(t') ,
\end{equation}
which leads to the following local equation of motion for the Nambu-Goldstone
boson:\footnote{This approximation was questioned in Ref.~\cite{Cline:2025yvg}.
The delta-function replacement used in this intermediate step is indeed not a valid
identity and should not be taken as the basis for extracting the local
coefficient. The local dissipative equation used below can instead be motivated
by an adiabatic/Markovian reduction of the nonlocal memory term, provided the
background varies slowly over the microscopic memory time. In the present
model, the corresponding adiabaticity conditions are parametrically satisfied.
Since $\Gamma$ is treated as a phenomenological parameter in the numerical
analysis, the results do not rely on the microscopic estimate
$\Gamma_{\rm eff}=g^2\omega/\pi$ obtained from the delta-function replacement.
The adiabatic/Markovian treatment is discussed in detail in the revised version
of our original article~\cite{Krasnov:2025fbo}.}
\begin{equation}
\Ddot{\theta} +\Gamma_{\mathrm{eff}}\Dot{\theta}+\cfrac{\Lambda^4}{f^2}\sin{\theta}=0,
\end{equation}
where $\Gamma_{\mathrm{eff}}$  is the effective damping rate.

To solve this equation, we rewrite it using dimensionless variables (where the prime denotes a derivative with respect to $\Lambda^2t/f$):
\begin{equation}\label{Eq_dimless_mink}
{\theta''}+\Gamma{\theta'}+\sin{\theta} = 0,
\end{equation}
where the dimensionless damping rate is defined as
\begin{equation}
\label{dimensionlessGamma}
\Gamma=\Gamma_{\rm eff}f/\Lambda^2.
\end{equation} 
We treat $\Gamma$ as a free phenomenological parameter in our calculations and explore both small ($\Gamma\leq1$) and large ($\Gamma > 1$) values of $\Gamma$.

\begin{figure}[H]
\centering 
\subfigure[Numerical solutions for sample values of $\Gamma\leqslant1$.]{\label{fig:a}\includegraphics[width=0.6\textwidth]{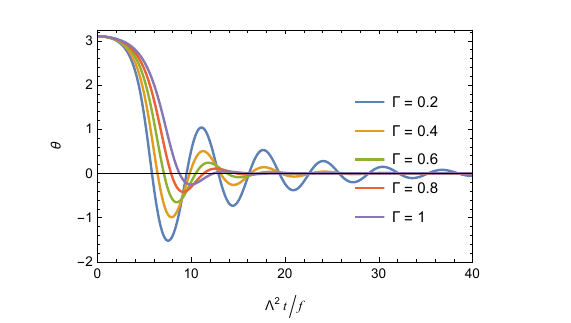}}
\subfigure[Numerical solutions for sample values of $\Gamma> 1$.]{\label{fig:b}\includegraphics[width=0.6\textwidth]{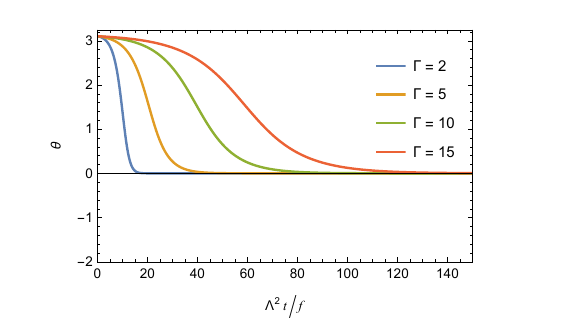}}
\caption{Numerical solutions of Eq.~\eqref{Eq_dimless_mink} with initial conditions $\theta_{in}=3.1$ and $\Dot{\theta}_{in} = 0$ for different values of $\Gamma$ in Minkowski space.} \label{NumSol0}
\end{figure}

Figure~\ref{NumSol0} displays numerical solutions to Eq.~\eqref{Eq_dimless_mink} for different $\Gamma$ values, starting from an initial phase near $\pi$. {\color{black} The results are shown in two subfigures for clarity. Unlike the case of small oscillations, we observe that larger $\Gamma$ values result in a longer duration for the field to reach the potential minimum.}
{\color{black} This behavior stems from the large initial phase, which causes the potential term in the equation of motion to behave differently compared to the small oscillation regime.}

\section{Baryon Asymmetry Calculation\label{baryonasymmetry}}
This section presents the calculation of the baryon asymmetry using the solutions to the equation of motion obtained previously.

Following \cite{Dolgov_1997}, the baryon ($B$) and antibaryon ($\overline{B}$) number densities in Minkowski space are given by:
\begin{equation}\label{BarDensity}
n_{B,\overline{B}}=\cfrac{g^2f^2}{2\pi^2}\int_0^\infty \omega^2 d\omega \left|\int_{-\infty}^{+\infty} e^{2i\omega t \pm i\theta(t)}dt \right|^2,
\end{equation}
where $+\theta(t)$ corresponds to baryons and $-\theta(t)$ to antibaryons. Note that $\omega$ in these integrals is not the same variable as in the equations of motion, despite the shared notation.

Defining the time integral as:
$$\int_{-\infty}^{+\infty} e^{2i\omega t \pm i\theta(t)}dt = N_\pm(\omega)\;,$$ 
it can be shown that:

\begin{equation}\label{N}
N_\pm(\omega) = -\cfrac{ie^{\pm i\theta_i}}{2\omega}+\cfrac{i}{2\omega}+\int_{0}^{+\infty} e^{2i\omega t }(e^{\pm i\theta(t)}-1)dt,
\end{equation}
where we omit delta functions due to the $\omega^2$ factor in the outer integral. This is further justified by the strict lower limit $\omega = m_Q+m_L>0$. The final term in \eqref{N} is evaluated numerically, similar to the integral in the previous section.

\begin{figure}[H]
\centering 
\subfigure [Numerical solutions for sample values of $\Gamma\leqslant1$.]{\label{fig:a3}\includegraphics[width=0.6\textwidth]{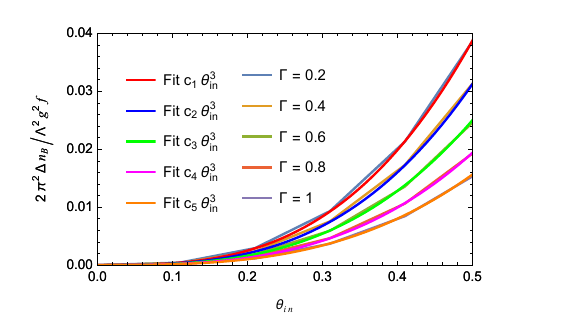}}
\subfigure[Numerical solutions for sample values of $\Gamma> 1$.]{\label{fig:b3}\includegraphics[width=0.6\textwidth]{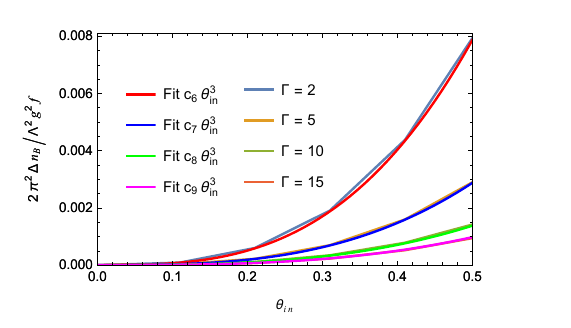}}
\caption{Baryon asymmetry in Minkowski space for a small initial phase and larger $\Gamma$ values, with cubic fit functions. This serves to validate our methodology. The coefficients $c_i$ are: $c_1\approx0.31,,c_2\approx 0.25,,c_3\approx 0.2,,c_4\approx 0.155,,c_5\approx 0.125, c_6\approx0.063,,c_7\approx 0.023,,c_8\approx 0.011,,c_9\approx 0.0078$. } \label{Res1}
\end{figure}

We now proceed to calculate the baryon asymmetry. First, we verify that our method reproduces the results of Ref.~\cite{Dolgov_1997}, {\color{black} where the baryon asymmetry was found to scale as $\theta^3_{in}$ for small oscillations}. For this purpose, we consider small initial phase values and plot the results with a cubic fit, as shown in Fig.~\ref{Res1}.

\begin{figure}[H]
\centering 
\subfigure[Numerical solutions for sample values of $\Gamma\leqslant1$.]{\label{fig:a4}\includegraphics[width=0.6\textwidth]{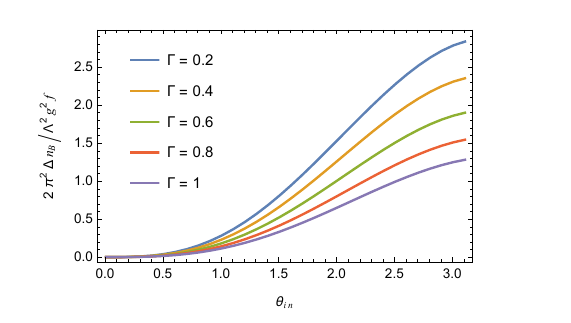}}
\subfigure[Numerical solutions for sample values of $\Gamma> 1$.]{\label{fig:b4}\includegraphics[width=0.6\textwidth]{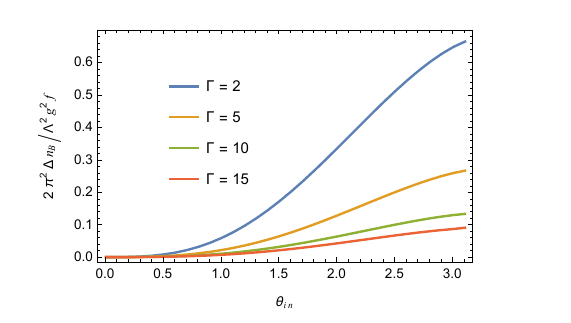}}
\caption{Baryon asymmetry $\Delta n_{B}$ as a function of the initial phase in Minkowski space-time. Particle production increases rapidly until $\theta_i \approx 1$, after which the rate decelerates considerably, tending toward saturation as $\theta_i$ approaches $\pi$. The curve's behavior is not strongly influenced by the value of $\Gamma$ when it is small. {\color{black} However, for larger $\Gamma$ values (as seen in Fig.~\ref{fig:b4}), the effects are more pronounced.} } \label{Res2}
\end{figure}

Next, we present the results for larger initial phases. The baryon asymmetry in Minkowski space is displayed in Fig.~\ref{Res2}. For small oscillations, \textcolor{black}{the oscillation period} is $T\sim 1/m_\theta$, but this relation does not hold for a large initial phase. The apparent saturation of particle production as the initial phase approaches $\pi$ is likely due to the oscillation period becoming significantly longer than the harmonic approximation would suggest. Although a deviation from the cubic dependence is evident, the calculated values remain of the same order of magnitude.

\section{Discussion and Conclusion}
We have re-examined the spontaneous baryogenesis scenario mediated by a Nambu-Goldstone boson. The common practice in the literature has been to employ the small-angle approximation for the cosine potential. However, the phase's probability distribution, shaped by quantum fluctuations during inflation, implies a non-negligible likelihood for substantial phase variations. This calls into question the reliability of the small-angle approximation and underscores the need to study large misalignment angles.

The primary aim of this paper was to investigate the key consequences of deviating from the small-angle approximation. As a first step, we worked within Minkowski spacetime, neglecting universe expansion, which is valid when the \textcolor{black}{decay rate $\Gamma$ of the pNGB field oscillations} is significantly greater than the Hubble expansion rate.

We computed the baryon asymmetry for an initial phase near $\pi$, as this value, located at a local maximum of the cosine potential, represents the most extreme case of large misalignment.
Our analysis, illustrated in Fig.~\ref{Res2}, reveals that the effects of a large misalignment angle on the generated baryon asymmetry are not substantially different from those predicted by the small-angle approximation in Minkowski space.

\section*{Acknowledgements}
The work of M. K. was performed in Southern Federal University with financial support of grant of Russian Science Foundation № 25-07-IF


\end{document}